\documentclass[12pt, letterpaper]{article}
\usepackage[utf8]{inputenc}
\usepackage{amsmath}
\usepackage{graphicx}
\usepackage{authblk}
\graphicspath{ {./images/} }
\usepackage[utf8]{inputenc}
\usepackage{amssymb}
\usepackage{multicol}
\usepackage{dcolumn}
\usepackage{changepage}
\begin{document}



\title{Recent Developments in the Penrose Conjecture}
\author{Hollis Williams }

\affil[$1$]{
University of Warwick,

Coventry, CV4 7AL, UK}

\maketitle

\begin{abstract}
We survey recent developments towards a proof of the Penrose conjecture and results on Penrose-type and other geometric inequalities for quasi-local masses in general relativity.

\end{abstract}

\section{Introduction}
\label{}
\noindent
The Penrose inequality relates the total mass of a black hole to its area.  To date, there is not a general proof for this inequality, although some important special cases have been proved.  In this article, we will present some recent important developments towards a proof of the Penrose conjecture as well as a range of recent results which have been obtained on Penrose-like and other geometric inequalities for quasi-local masses.  In Section 2, we will outline the necessary background and preliminaries (mostly for the benefit of mathematical readers who have not previously studied general relativity).  In Section 3, we sketch the Penrose conjecture and some of the well-known partial results.  In Section 4, we collect some recent results on the Penrose inequality for perturbations of Schwarzchild spacetime, spinor approaches to the Penrose inequality, counterexamples to Penrose-type inequalities and cosmic censorship, and results on extended Penrose-type inequalities using other types of quasi-local mass apart from the ADM mass.

\section{Preliminaries}
\noindent
In this section we will present some well-known material on Lorentzian geometry, the Einstein equations and the positive mass theorem which is mostly intended for beginners to general relativity.  General relativity represents a unique intersection point of geometry, analysis and mathematical physics, perhaps the only intersection where the contribution from both sides from mathematics and physics has been equal in terms of intuition and rigour and certainly the only current intersection of geometry and physics which connects with and approximates reality in a way which is verified experimentally.  More recently, there have also been some very exciting developments in the intersection between geometry and string theory (for example, mirror symmetry and enumerative geometry).  

\subsection{Lorentzian Geometry}

\noindent There are many definitions in Lorentzian geometry which take some time to digest and we will mention a few of these, but there is a relatively small number of core theorems [1,2].  A Lorentzian manifold is an an $(n+1)$-dimensional pseudo-Riemannianian manifold $(N,g)$ such that at every $p$ in the manifold, the inner product $g$ defined as 

\[g: T_{p}M \times T_{p}M \rightarrow \mathbb{R} \tag{1}\]

\noindent
has signature $(-,+,...,+)$ referred to as Lorentz signature.  The pseudo-Riemannian case is in direct contrast to Riemannian manifolds where the metric is positive definite.  Results about curvature in Riemannian geometry typically carry over to the Lorentzian case but there are many theorems which do not transfer to pseudo-Riemannian geometry [3].  As an example of such a result, the restriction of a pseudo-Riemannian metric to a submanifold may not be non-degenerate even though the metric on the ambient manifold is.  If we take two vectors $X$ and $Y$ and the Minkowski metric $\eta$, we obtain

\[\eta(X,Y) = -X^0 Y^0 + \sum_{i=1}^n X^i Y^i, \tag{2} \]
\noindent
This implies that tangent spaces to points on a Lorentzian manifold are isometric to Minkowski space, so general relativity 'looks like' special relativity if we zoom in close enough, exactly analogous to the way that a general curved manifold 'looks flat' if we zoom in close enough.  

The Lorentz signature allows us to classify tangent vectors into three different types.  A vector $X$ is timelike if $g(X,X) < 0$, spacelike if $g(X,X) > 0$ and null if $g(X,X)=0$.  A tangent vector $X$ is said to be causal if it is either timelike or null.  This terminology may be familiar from a second-year geometry course, but in Lorentzian geometry it takes on a physical significance suggested by the adjective 'causal'.  The set of null vectors in the tangent space at a point $p$ forms a double cone in that space known as a light cone.  An event on (or in) the lightcone is
a past or future event which could be causally related to the
event at the origin.  Timelike vectors point inside the light cone and null vectors lie on it, but they both represent events which are causally related, whereas as spacelike vectors point outside the double cone and do not reprsent causally related events.   Physically, the light cone at an event can be viewed as set of points which can be linked to that event by light signals.  Time-like or null vectors in the future cone are said to be future-pointing, whilst those in the past cone are past-pointing.  An alternative way of phrasing this is to split the time-like vectors into those which have a positive and those which have a negative first component and to say that the former are future-directed and that the latter are past-directed.

However, things are not quite that simple because there is a corresponding notion of orientability on Lorentzian manifolds known as 'time orientability'.  When we consider a light cone in a tangent space $T_p M$, we must make a choice as to which is the future cone and which is the past cone.  A manifold $N$ is time-orientable if one can make a continuous designation of future and past cones as one moves through all the tangent spaces across the manifold.  A time orientation can be fixed on a time-orientable manifold by making a choice for a smooth time-like vector field without zeroes.  A causal non-spacelike vector $X$ is then future pointing if it points into the same cone as the chosen vector field.  This notion of time-orientation allows us to make precise the idea of a spacetime: a spacetime is a connected (not necessarily simply connected), time-oriented Lorentzian manifold $(N,g)$.  The exact definition may vary slightly from one text to another.

This notion of causality can be extended easily to smooth curves $\gamma : I \rightarrow N$.  A curve is said to be timelike if the tangent vector $\gamma'(t)$ is timelike for all $t \in I$, with the analogous definition for null and spacelike curves.  As before, a non-spacelike curve is said to be causal.  If $\gamma$ is a timelike curve, the arc length in the corresponding arc length parametrisation of the curve corresponds to proper time (the time measured by a clock carried by an observer as they travel along a world line).  Proper time is generally not the same as coordinate time.  When working with the Schwarzchild metric, it so happens that the difference in Schwarzchild coordinate time $d t$ between two events is equal to the proper time difference $d \tau_{\infty}$ between the events as measured by a stationary observer at spatial infinity, but this is unusual.  The arc length function for an admissible causal curve $\gamma: [a,b] \rightarrow \mathbb{R}$ is defined as 

\[L(\gamma) = \int_a^b |\gamma' (t)| \: dt. \tag{3} \]

\noindent
There is also the corresponding notion of a future-directed and a past-directed causal curve defined in terms of the tangent vectors to points on the curve.  To give a bit more of an idea about the causal relationships between points which one can study on a Lorentzian manifold, note that one can define both a timelike and a causal future (respectively, past) of a point $p$ in a spacetime via the following sets:

\[I^{+}(p) = (q \in N : p \ll q), \: \: J^{+}(p) = (q \in N : p \leq q), \tag{4} \]

\noindent
where the first causal relation means that there exists at least one future directed timelike curve connecting $p$ to $q$ and the second relation means that there is a corresponding future directed causal curve.  We cannot say much more about Lorentzian geometry here and refer the reader to the lecture notes of Galloway, but we should mention that the notion of causality can be generalised from curves to some submanifolds of higher dimension (ie. hypersurfaces).  A spacelike hypersurface is a hypersurface whose tangent vectors are all spacelike.  In geometric terms, the induced metric from the ambient manifold is Riemannian.  The intuitive interpretation of a spacelike hypersurface is that represents the 'space' part of a spacetime when time is stopped for an instant.  

There are various theorems in general relativity which are distinguished by their geometric beauty as well as their physical significance.  A notable example is Hawking's black hole topology theorem, which states that cross sections of the event horizon in an asymptotically flat stationary black hole spacetime which obeys the dominant energy condition have the topology of the sphere (the Euler characteristic of the cross sections is $2$).  Hawking also extended this result to outer apparent horizons in non-stationary black hole spacetimes and Schoen and Galloway generalized Hawking's theorem to a statement about the topology of black holes in higher dimension [4].  Extension of classical results in general relativity to higher dimensions is clearly important if we hope to have higher-dimensional theories of gravity, although the extended result is generally not the exact equivalent of the classical statement.  As an example, it is known that the Einstein field equations are the unique second-order, quasilinear PDEs for the coefficients of the metric tensor $g_{i j}$ modulo addition of a cosmological constant, but only in dimenson $4$ [5].  These equations are the topic of our next section.

\subsection{The Einstein Field Equations}

\noindent 
The physical motivation behind general relativity is to explain gravity as a consequence of the curvature of spacetime caused by the presence of matter and radiation.  Minkowski realised that special relativity (the version of the theory which has only inertial and no gravitational forces) can be formulated on a Lorentzian manifold called Minkowski spacetime (the Lorentzian analogue of flat Euclidean space).  Since Minkowski spacetime is both empty and flat, Einstein had the idea that one could instead take a general Lorentzian manifold $(N^4,g)$ and link the energy in the spacetime with its metric $g$.  The connection coefficients $\Gamma_{ij}^k$ of the metric on a Lorentzian manifold represent both inertial and gravitational forces by the principle of equivalence of gravitation and inertia, whereas only inertial forces appear through the connection coefficients in the analogous dynamical equations on Minkowski spacetime.

The next step is to find equations which link the curvature of the metric $g$ with its matter (or rather, energy) content.  We would like to write our physical quantities as tensor fields because they are intrinsic geometric objects with well-understood transformation laws, so the natural thing to do would be to take the stress-energy tensor $T$ and set it equal to some curvature tensor.  The stress-energy tensor can be defined to be the multi-linear object such that $T(v,w)$ is the energy density in the direction of a future-directed unit time-like vector $v$ as measured by an observer who is travelling in the direction of another future-directed unit time-like vector $w$, where $v$ and $w$ are both tangent vectors at some point $p$ in $N$.  

The physical observation that all energy densities are non-negative translates into a simple mathematical statement known as the dominant energy condition:

\[T(v,w) \geq 0. \tag{5}\]

\noindent
The most obvious choice of curvature tensor which we can set equal to $T$ is the Ricci tensor $\text{Ric}$, since this is also a $2$-tensor.  However, a computation shows that 

\[\text{div}(\text{Ric}) \neq 0. \tag{6}\]

\noindent
One would expect conservation of energy and momentum in a reasonable physical system, which corresponds to the equation 

\[\text{div}\: T = 0. \tag{7} \]

Although the divergence of $\text{Ric}$ is non-zero, we can instead define the Einstein tensor

\[G = \text{Ric} - \frac{1}{2} R g,  \tag{8}\]

\noindent
where
\[\text{div} \: G = 0 . \tag{9}\]

\noindent
The conservation identity can be proved with a computation using the  differential Bianchi identity for the covariant derivative of the Riemann tensor.  The computation is straightforward but there seem to be some incorrect proofs in several popular relativity textbooks, so we will state the correct computation.  Beginning with the differential Bianchi identity, we have

\[ \nabla_a R_{b c d e} + \nabla_e R_{bcad} + \nabla_d R_{b c e a} =0, \tag{10a}\]

\[\nabla_a g^{ia} g ^{bd} g^{ce} R_{b c d e}  + \nabla_e g^{ia} g ^{bd} g^{ce} R_{bcad} + \nabla_d  g^{ia} g ^{bd} g^{ce}R_{b c e a}.\tag{10b} \]

\noindent
Contracting and using symmetries of the Riemann tensor, we obtain

\[ \nabla_a g^{ia} R - \nabla_e g^{ia} g^{bd} g^{ce} R_{d a bc} - \nabla_d g^{ia} g^{bd} g^{ce} R_{e a c b} = 0. \tag{11}\]

\noindent
Contracting further and re-arranging, we get

\[ \nabla_a g^{ia} R - \nabla_e g^{i a} g^{c e} R_{a c} - \nabla_d g^{i a} g^{b d} R_{a b} =0, \tag{12a}\]

\[\nabla_a g^{i a} R - \nabla_e R^{i e} - \nabla_d R^{i d} =0, \tag{12b}\]

\[\nabla_a g^{i a} R - 2 \nabla_a R^{i a} =0, \tag{12c}\]

\[-2\nabla_a \bigg(R^{i a} - \frac{1}{2} g^{i a} R \bigg) =0. \tag{12d}\]

\noindent
This implies the result.

It was this conservation identity and analogy with the classical Poisson equation that led Einstein to propose the Einstein field equations:

\[G = 8 \pi T. \tag{13}\]
\noindent
Since $\text{div} \: G =0$, we get the required conservation property which $T$ must satisfy in order to be compatible with the field equations.  For this reason it is sometimes said that the twice contracted second Bianchi inequality implies the Einstein equations.  To call this a derivation of the equations is perhaps a bit of a hand wave, but the argument is one that appeals to geometers and it is close to Einstein's reasoning in his original derivation.  The 'proper' derivation (especially from the viewpoint of modern theoretical physics) is to use the Einstein-Hilbert action (defined to be the total integral of the scalar curvature of the metric).

There are two main problems with these equations: firstly, when written out fully in local coordinates, they form a system of $10$ non-linear PDEs with both elliptic and hyperbolic properties and secondly, solving them would involve already having complete knowledge of the stress-energy tensor field (not necessarily a symmetric tensor).  The latter is unlikely unless we prescribe it to have some particularly simple form.  A perfect fluid, for example, has the following stress-energy tensor:

\[T_{i j} = \mu u_i u_j + p(g_{i j} + u_i u_j), \tag{14}\]

\noindent
where $u$ is the velocity and $\mu$ and $p$ are the energy and pressure densities.  The first step which Einstein took when solving the equations is to make approximations and this is still perhaps the main method of solution, forming a branch of the subject called numerical relativity.  Computers are certainly not infallible and it is worth pointing out that conjectures have in the past been assumed in numerical relativity which were then disproved (see Thorne's hoop conjecture for an example).

A solution to the second problem would be to simplify the equations considerably by assuming that there are no matter sources in the spacetime and setting $T$ equal to $0$.  This would seem to imply that

\[G =0,\]

\noindent
but we can say more than that by tracing both sides of the equation.

\[\text{Ric} =  \frac{1}{2} R g,\tag{15a}\]
\[R =  2 R. \tag{15b}\]

\noindent
Recall from linear algebra that the trace is invariant with respect to a change of basis, so the trace of the metric with respect to the metric is just the dimension of the manifold.  This can only be satisfied if the scalar curvature is zero, so the vacuum Einstein equations are in fact

\[\text{Ric} = 0. \tag{16}\]

\noindent
These equations are the Euler-Lagrange equations of the Einstein-Hilbert action: the easiest way to see this is to take the action and vary it.

\[\mathcal{L} = \int R \: \mu_g, \tag{17}\]

\noindent
where $\mu$ is the Minkowski volume element.  The full equations are obtained by adding a term to the Lagrangian which describes the matter fields and varying as before.  

Geometrically the equations tell us that vacuum spacetimes (Schwarzchild, for example) are Ricci flat, but if we want to work with PDEs we will need to write the equations out in their local coordinate representations in charts whose domains are included in an open set $U$.  If we do this with the Ricci tensor, we obtain an explicit system of second-order hyper-quasilinear PDEs for the coefficients of the metric:

\[R_{i j} = \frac{\partial}{\partial x^k} \Gamma_{i j}^k - \frac{\partial}{\partial x^i} \Gamma_{j k}^k + \Gamma_{ ij}^k \Gamma_{k d}^d - \Gamma_{i d}^k \Gamma_{j k}^d, \tag{17}\]

\noindent
where the connection coefficients $\Gamma_{i j}^k$ are the $n^3$ functions which describe a linear connection in local coordinates with respect to a frame:

\[\Gamma_{\alpha \beta}^{\lambda} = \frac{1}{2}g^{\lambda \mu} \bigg( \frac{\partial g_{\alpha \mu}}{\partial x^{\beta}} + \frac{\partial g_{\beta \mu}}{\partial x^{\alpha}}  - \frac{\partial g_{\alpha \beta}}{\partial x^{\mu}} \bigg). \tag{18} \]

\noindent
To clarify, a PDE is second-order quasilinear if it is linear in its second-order derivatives and hyper-quasilinear if the coefficients of the second-order derivatives do not contain first-order derivatives.

The first trivial example of a vacuum spacetime is Minkowski spacetime: this has vanishing Riemann tensor and hence a vanishing Ricci tensor.  

\[(\mathbb{R}^4, -dt^2 + d x^2 + dy^2 +d z^2). \tag{19}\]

\noindent
The next simplest family of solutions is given by the Schwarzchild spacetime:

\[ \bigg(\mathbb{R} \times (\mathbb{R}^3 \ B_{m/2}(0)), - \bigg( \frac{1 - \frac{m}{2r}}{1 + \frac{m}{2r}} \bigg)^2 dt^2 + \bigg( 1 + \frac{m}{2r} \bigg)^4  (dx^2 + dy^2 + dz^2). \tag{20}\]

\noindent
This is one of the most important examples of a warped product metric: that is, a metric of the form

\[G = dr^2 + \rho^2(r)g_M, \tag{21}\]

\noindent
where $g_M$ is a Riemannian metric on a manifold $M$ and $\rho$ is strictly positive on an open interval $I$: the warped product is then a metric on the product manifold $M \times I$.  In physical terms, these spacetimes represent static black holes.  Another exact solution is the FLRW metric used in cosmology to represent a cosmological model of a homogeneous, isotropic universe.  Combining the Robertson-Walker metric with the stress-energy tensor for an ideal fluid allows one to write the Einstein equations in the form of two ordinary differential equations known as the Fridman equations.  

\[H^2 = \frac{8 \pi}{3} \rho - \frac{k c^2}{a^2},\tag{22a} \]

\[\dot{H} + H^2 = -\frac{4 \pi}{3} \bigg(\rho + \frac{3 p}{c^2} \bigg). \tag{22b}\]

\noindent
These equations relate the scale factor $a$ of the cosmological model, the curvature parameter $k$ and the densities and pressures of the fluids.  $H$ denotes the Hubble parameter, which quantifies the rate at which the universe is increasing in size due to expansion.    

For one final exact solution, we mention that one can obtain a metric which describes a rotating black hole in a vacuum spacetime known as the Kerr metric, but the derivation is very involved.   Although the Kerr metric is stationary state, it is not static as the black hole (or the stationarity Killing vector) is allowed to rotate.  A metric  is static if it does not contain any cross terms such as $dr \: dt$.  Sending $t$ to $-t$ will obviously have no effect on such a metric, so a static metric is automatically time-symmetric.  The converse is not true, however: one can have FLRW spacetimes in which there is a clear notion of time reversal or some notion of a fundamental observer who measures a privileged proper time, and yet the spacetime is clearly not static.  It was proved by Bunting and Masood-ul-Alam that Minkowski and Schwarzchild spacetime are the only two possibilities for complete, asymptotically flat static vacuum spacetimes [6].  We will come to the definition of asymptotically flat shortly.  

\subsection{The Positive Mass Theorem}

\noindent
There are many physical theorems in general relativity: once formulated in a geometric way these become theorems of Lorentzian differential geometry.  As an example, consider the positive energy theorem (alternatively known as the positive mass theorem).  The physical problem is that there is no satisfactory way of describing the local energy density of a gravitational field, so one has to define the total energy of an isolated gravitating system in terms of the asymptotic behaviour of the field as one goes out to infinity.  The positive energy theorem states that the total energy defined in this way is zero in the case of Minkowski spacetime and positive in all other cases.  The exact definition of the total energy is as a limit of surface integrals over spheres in the asymptotically flat region of the spacetime:

\[m(g):= \lim_{r \rightarrow \infty} \int_{S_{r}} \partial_j g_{ij} - \partial_i g_{j j} \: dA^i.\tag{23}\]

\noindent
An equivalent way of viewing the problem is that a gravitating system with non-negative local mass density must have non-negative total mass when measured at spatial infinity.  The ADM mass is then defined in an equivalent way.  The rigorous geometric formulation is as follows: start with a triple $(M,g,k)$ (known as a Cauchy data or an initial data set in the literature), where $k$ is a symmetric tensor field, $M$ is a Riemannian $3$-manifold and where $g$ and $k$ satisfy equations of constraint for the local matter and momentum densities:

\[16 \pi \mu = R - k^{ij}k_{ij} + (g^{ij}k_{ij})^2, \tag{24a}\]

\[8 \pi J_i = \nabla^j (k_{ij} - (g^{ab} k_{ab}) g_{ij}.\tag{24b}\]

\noindent
These correspond to the Gauss-Codazzi equations from differential geometry.

If this initial data set is asymptotically flat and satisfies the dominant energy condition, then the mass is non-negative and vanishes only in the case of Minkowski spacetime.  More precisely, the mass is zero if the data is the pullback of the Cauchy data induced on the image of a space-like embedding of $M$ into Minkowski space.  The dominant energy condition is written here as

\[ \mu \geq |J|.\tag{25}\]

\noindent
A Cauchy data is asymptotically flat if for some compact set $C$, $M \setminus C$ consists of a finite number of components $N_k$ (known in the literature as 'ends') such that $N_k$ is diffeomorphic to the complement of a compact set in $\mathbb{R}^3$.  Under these diffeomorphisms, the metric is written as

\[ds^2 = \bigg(1 + \frac{m}{2r} \bigg)^4 \sum_i (dx^i)^2 + \sum_{i,j} p_{ij} dx^{i} dx^{j}, \tag{26}\]

\noindent
where
\[p_{ij}= O \bigg(\frac{1}{r^2} \bigg), \: \: \nabla p_{ij} = O \bigg(\frac{1}{r^3} \bigg), \: \:   \nabla \nabla p_{ij}= O\bigg(\frac{1}{r^4} \bigg).  \tag{27}\]

\noindent
The quantity $m_i$ the ADM mass of the end $N_i$.  The theorem then says that for an asymptotically flat initial data set, each end has non-negative total mass and if one of the ends has zero mass, then the initial data set vanishes and the Riemann tensor is zero.

Rather than attempting to prove the general result, one can assume that the $3$-manifold is a maximal spacelike hypersurface of the spacetime in which it is embedded.  This corresponds to a vanishing trace of the second fundamental form, so we now have $k=0$.  In the geometric formulation, the dominant energy condition becomes the condition that $M$ has non-negative scalar curvature.  This special case is known as the 'time-symmetric' problem, since it is equivalent to the hypersurface being time-symmetric or totally geodesic, where a totally geodesic Riemannian submanifold is one such that for every $V$ in the tangent bundle, the geodesic $\gamma_V$ lies entirely in the larger manifold when viewed in terms of the induced metric on the submanifold.  Schoen and Yau proved the time-symmetric case in 1979 with an argument by contradiction whereby one assumes the total mass of an end to be negative and uses this to derive the existence of a complete area minimizing surface [7].  The surface is then shown not to exist when the scalar curvature is non-negative via a second variation inequality for the stability of minimal surfaces.

In 1981, Schoen and Yau showed that the positive mass theorem for general data can always be reduced to the time-symmetric case using a quasilinear elliptic PDE called the Jang equation (introduced by the physicist Jang in 1978 for this purpose) [8].  This is done by searching for a hypersurface $\Sigma$ in $M \times \mathbb{R}$ called the Jang surface with scalar curvature which is as positive as possible.  The Jang surface is required to satisfy the Jang equation:

\[H_{\Sigma} = \text{Tr}_{\Sigma} k . \tag{28}\]

\noindent
In words, the mean curvature at a point is prescribed to equal the trace of the trivial extension of $k$ when it is restricted to the Jang surface.  If we consider the hypersurface to be given by a graph $t=f(x)$ in the product space $M \times \mathbb{R}$ with the appropriate product metric, the Jang equation takes the following form in local coordinates:

\[ \bigg( g^{ij} - \frac{f^i f^j}{1 + |\nabla f|^2} \bigg) \bigg( \frac{\nabla_{ij}f}{\sqrt{1 + |\nabla f|^2}} - k_{ij}\bigg) =0. \tag{29}  \]

\noindent
Some work then has to be done to show that this equation has the necessary existence and regularity results: this is done using estimates from the theory of elliptic PDEs.  In 1981 Witten outlined an alternative proof using identities for Dirac operators on spinors [9].  

It is not currently clear if the positive mass theorem holds in arbitrary dimension.  Witten's proof assumes that the manifolds are spin manifolds and the proof by Schoen and Yau only holds for dimensions between $3$ and $7$ because the relevant regularity theory for minimal hypersurfaces is only valid up to dimension $7$.  Lohkamp claims to have a singularity excision argument which proves the theorem in any dimension, but his proof has not been published or verified.  Schoen and Yau have also claimed that they have been able to extend their minimal surface techniques for positive scalar curvature problems to dimensions beyond $7$.  If true, the positive mass theorem in arbitrary dimension would follow roughly as a corollary.

\section{The Penrose Conjecture}
\noindent
We move on to a conjecture which can be viewed as a sharpening of the positive mass theorem to spacetimes which contain black holes.  The conjecture states that the following inequality called the Penrose inequality holds between the mass of a spacetime and the total area of the black holes which it contains:

\[M \geq \sqrt{\frac{A}{16 \pi}}. \tag{30}\]

\noindent
Note that the spacetime must be asymptotically flat in order to have a definition of the ADM mass.  One can consider $A$ to be the area of a spacelike cross section of the horizon when there is one black hole, or the sum of such areas if there is more than one black hole.  $A$ will generally depend on time-slicing, hence why the Penrose inequality is often given as a conjectured property of an asymptotically flat Cauchy surface.

Geometric analysts might like to view the inequality as a version of the isoperimetric inequality for black holes.  The inequality was proposed by Penrose using a physical argument which incorporates many of the classic assumptions which are taken for granted in general relativity (the cosmic censorship hypothesis, for example) [10].  A counterexample to the conjecture would offer a serious challenge to the establishment view of general relativity and might suggest that naked singularities exist and can be observed in the universe outside of black holes.  The experimental evidence which we have from violent black hole mergers suggests that this very likely cannot happen.

We know that there is a time-symmetric and a general version of the positive mass theorem and the analogous cases also exist for the Penrose inequality.  If we may take a brief digression for some definitions, define $S$ to be the collection of surfaces which are smooth compact boundaries of open sets $U$ in a $3$-manifold $M$, where $U$ contains all the points at infinity which are added in to compactify all of the ends $N_k$.  Given a surface $\Sigma \in S$, define $\bar{\Sigma} \in S$ to be the outermost minimal area enclosure of $\Sigma$.  'Outermost minimal' simply refers to a minimal surface which is not contained completely within another minimal surface.  Outermost minimal surfaces are always spheres, so one may see them referred to as outermost minimal spheres.  For a geometric picture of the outermost minimal area enclosure, think of a simple $3$-manifold like $B^3$ and then visualise $\mathbb{R}^3 \setminus B^3$.  The boundary of the ball can be viewed as the horizon of the black hole and the OMAE is a surface which lives in $\mathbb{R}^3 \setminus B^3$ and wants to get as close as possible to $B^3$.  In the case of the Euclidean metric, the OMAE coincides with the boundary of the ball.  Finally, define $\Sigma \in S$ in $(M^3,g,k)$ to be a future apparent horizon if

\[H_{\Sigma} + \text{Tr}_{\Sigma} k =0. \tag{31}\]

To return to the inequalities, the Riemannian Penrose inequality states that if we have a asymptotically flat Riemannian $3$-manifold with vanishing second fundamental form and non-negative scalar curvature and if $\Sigma \in S$ is a minimal surface, then the Penrose inequality is satisfied when $A$ is the outermost minimal area enclosure of $\Sigma$.  Furthermore, equality only occurs if $(M^3 \setminus U,g)$ is isometric to Schwarzschild spacetime.  Notice that we are now reducing back to Schwarzchild space rather than Minkowski space, since Minkowski space does not contain horizons.  For the general Penrose inequality, we start with a Cauchy data, the dominant energy condition and an asymptotically flat $3$-manifold, but now, if $\Sigma \in S$ is a future apparent horizon, then the Penrose inequality is satisfied when $A$ is the outermost minimal area enclosure of $\Sigma$.  Equality only occurs if the Cauchy data is the pullback of the data induced on the image of a space-like embedding of $M^3 \setminus U$ into the exterior region of Schwarzchild spacetime.

In 2001 Huisken and Ilmanen published a proof of the Riemannian Penrose equality using inverse mean curvature flow (strictly speaking, they proved that the total mass of a spacetime is related to the area of the largest black hole contained in the spacetime) [11].  A map $F$ from a family of hypersurfaces $M^n \times [0,T) $ to $(N^{n+1}, \bar{g}$ solves mean curvature flow if 

\[\frac{d}{dt} F(p,t) = \vec{H}(F(p,t)) ,\tag{32} \]

\noindent
where $\vec{H}$ is the mean curvature vector.  Written out in local coordinates, one sees that this is a quasilinear, second-order system of parabolic PDEs.  In more geometric terms familiar from geometric analysis, the flow evolves a family of surfaces in a Riemannian manifold by evolving the normal components of vectors at points on the surface by a speed which is prescribed to be equal to the mean curvature at that point. It is also possible to define a flow called inverse mean curvature flow, where the same is done for the inverse of the mean curvature. This would seem like a difficult flow to use in applications, since it can develop singularities easily and looks to be undefined if the mean curvature at a point is zero.  The idea of using an inverse version of mean curvature flow to prove the Riemannian Penrose inequality was first suggested by the physicist Geroch, but he was unable to overcome the problem that the flow can develop discontinuous jumps.

A proof of the inequality from this angle is technically very difficult as one needs an existence theory for the flow which allows for weak solutions.  More precisely, Huisken and Ilmanen used a level sets approach that the surface $\Sigma(t)$ which one obtains after flowing a minimal surface for a time $t$ can be defined as the level set of a function $u$ on $(M^3,g)$ such that

\[ \text{div} \bigg( \frac{\nabla u}{| \nabla u|} \bigg) = |\nabla u|.\tag{33} \]

\noindent
This is a degenerate elliptic PDE which can be shown to have a decent weak existence theory.  The Hawking mass is monotone on the level sets of $u$ and some further technical arguments show that it is eventually bounded over time by the ADM mass, which proves the result.  The positive mass theorem follows as a corollary.

In 2001, Bray published a proof of the Riemannian Penrose inequality in full generality using a different flow called conformal flow of metrics [12].  Flowing the metric under CFM gives a metric $g_t = u^4_t g$ at time $t$, where

\[\frac{d}{dt} u_t = v_t u_t, \tag{34a}\]  

\[\Delta_{g_{t}} v_t =0  \: \: \text{on} \: M_t, \tag{34b}\]

\[v_t = 0  \: \: \text{on} \: \partial M_t, \: \: v_t(x) \rightarrow -1, \: \: |x| \rightarrow \infty.\tag{34c} \]

\noindent
$M_t$ denotes $(M,g_t)$.  The basic argument is roughly as follows: the flow has been defined to stay within the conformal class of the original metric and to approach closer and closer to the Schwarzchild metric as the flow is continued for longer periods of time.  By construction the flow also keeps the area of the outermost minimal area enclosure of $\Sigma$ constant.  Since the total mass of $M^3$ can be shown to be non-increasing, one can apply the positive mass theorem after a reflection of the manifold through a minimal surface and a conformal compactification of one of the two resulting ends.  Since the Schwarzchild metric gives equality in the Penrose inequality, the inequality holds for the original manifold $M$.  This is a cartoon sketch of the idea of the proof, but again, the paper itself is very long and technical.

It remains an open problem to prove the Penrose inequality in full generality.  One possibility which crops up in the literature is construct a codimension-$2$ flow of $2$-manifolds within an arbitrary spacetime such that some quantity like the Hawking or Bartnik mass is non-negative along the flow and limits to the ADM mass.  Bray has suggested a flow of $2$-manifolds which he calls uniformly area-expanding time-flat flow [13].  It seems unlikely that such an approach could prove the full Penrose inequality, especially as one generally ends up with a flow which is a backwards heat equation and which consequently does not even have short-time existence.  It might be possible to get some important partial results using such a flow: for example, a proof in the case of perturbations away from Minkowski or Schwarzchild space.  Such a result would still be very far away from full generality, as one cannot obtain an arbitary manifold by perturbations of the metric because of topological obstruction.  It is not clear how one would generalise Witten's proof to the scenario of the Penrose inequality, so a proof from this angle of attack is also unlikely.

Another possibility would be to follow the example of Schoen and Yau and use a modified version of the Jang equation for Jang surfaces in a space with a warped product metric as opposed to a standard product.  If one considers a non-trivial extension $K$ of $k$ to all of $M^3 \times \mathbb{R}$ which allows solutions of the modified Jang equation to blow up at apparent horizons, then the Jang equation in local coordinates is not much more complicated

\[ \bigg( g^{ij} - \frac{ \phi^2 f^i f^j}{1 + \phi^2|\nabla f|^2} \bigg) \bigg( \frac{\phi \nabla_{ij}f + \phi_i f_j + \phi_j f_i}{\sqrt{1 + \phi^2|\nabla f|^2}} - k_{ij}\bigg) =0.  \tag{35} \]

\noindent
where $\phi$ is the warping factor of the Jang metric $\bar{g} = g + \phi^2 df^2$ on the Jang surface.  The blow up behaviour is desirable in this case as one wants to preserve the area of the horizon in the Jang surface.  However, the analysis and boundary behaviour is made much more difficult by the presence of the warping factor $\phi$, which depends on $f$ in a highly non-trivial way.  One has to couple the Jang equation to some set of equations which determines $\phi$: for example, we can couple the equation to one of the geometric flows which were used to prove the time-symmetric case.  The idea that the Penrose inequality could be proved by coupling a suitable flow to the generalized Jang equation was suggested and studied by Bray-Khuri [14,15].  The generalized Jang equation has been studied in the general case by Han-Khuri, with proofs of existence, regularity, and blow-up results [16].  However, the warping factor $\phi$ was prescribed in this analysis, and it is an open problem to try to get control of the quantities involved in the PDE without prescribing $\phi$.  This is to say nothing about getting the necessary existence and regularity result for the full Jang/CFM system: needless to say, this is a very tough open problem which would require some new technology to be resolved.  Han-Khuri later studied the Jang/CFM coupling in more detail and wrote down the system of equations explictly [17].  There are in fact two possibilities for the Jang/CFM system outlined in that paper (of which we will use the first).  In both cases, one starts with the generalized Jang equation

\[ H_{\Sigma} = \text{Tr}_{\Sigma} K. \tag{36}\]

\noindent
Along with this, one is also required to solve either the Dirac equation plus an equation for the warping factor,

\[ \mathcal{D} \tilde{\psi}_t =0, \tag{37a}\]

\[ \phi = 2 \int_0^{\infty} \chi_t ( (w_t^{-})^2 |\psi_t^{-}|^2 +  (w_t^{+})^2 |\psi_t^{+}|^2) u_t^2 \: dt ,\tag{37b}\]

\noindent
or a zero scalar curvature equation plus an equation for the warping factor

\[ \Delta_{\tilde{g}_t} z_t - \frac{1}{8} \tilde{R}_t z_t =0 \: \: \text{on} \:\: \tilde{M}_t, \: \: z_t \rightarrow 1 \: \text{as} \: \: |x| \rightarrow  \infty  , \tag{38a}\]

\[  \phi = 2 \int_0^{\infty} \chi_t ( ( w_t^{-} z_t^{-})^2 + ( w_t^{+} z_t^{+})^2) u_t^2 \: dt . \tag{38b}\]

\noindent
In both cases, one also has a boundary condition

\[\overline{H}_{\partial \Sigma_0} =0 , \tag{39}\]

\noindent
where $\overline{H}_{\partial \Sigma_0}$ is the mean curvature of the boundary of the Jang surface over the region outside the outermost minimal area enclosure of $\partial M$.

One may consult [17] for the derivation of these equations, but just to fix notation, let us say that $ \mathcal{D}$ is the Dirac operator and $\tilde{\psi}_t$ is a harmonic spinor on $\tilde{M}_t$ which converges to a constant spinor with unit norm at spatial infinity, where $\tilde{M}_t$ is the doubled manifold $ M_t^{-} \cup M_t^{+}$ reflected across a minimal surface $\partial M_t$.  The metric on $\tilde{M}_t$ is piecewise defined $\tilde{g}_t = g_t^{-} \: \cup \: g_t^{+}$ where $g_t^{\pm} = (w_t^{\pm})^4 g_t$, $w_t^{\pm} = (1 \pm v_t)/2$ and $\psi_t^{\pm}$ is the corresponding restriction of $\tilde{\psi}_t$.  $\tilde{R}_t$ is the corresponding scalar curvature of the metric as one might guess.  $\chi_t$ is an indicator function on $\Sigma_t$, where $\Sigma_t$ is the region outside the outermost minimal surface $\partial \Sigma_t$ ($\Sigma_0$ being a solution of the Jang equation whose metric is then evolved under conformal flow of metrics).  The conformal flow of metrics is given by $g_t = u_t^4 g$ where

\[ \frac{d}{dt} u_t = v_t u_t  , \tag{40a}\]

\[ \Delta_{g_t} v_t =0 \:\: \text{on} \:\: M_t ,\]

\[  \left.v_t\right|_{\partial M_t} = 0 , \:\: v_t(x) \rightarrow -1 \:\: \text{as} \:\: |x|  \: \rightarrow \infty, \tag{40b}\]

\noindent
where $M_t$ is the region outside of $\partial M_t$.  Note that any proof along these lines which uses the first system of equations is valid only for spin manifolds (for the same reason that Witten's proof only holds for spin manifolds), but this includes all the manifolds which are relevant for general relativity.  The Penrose inequality in higher dimensions is a separate matter.

\section{Recent Developments}

\noindent
So far what we have described is well-known.  This is a fast-moving subject, however, and there have been some significant developments in the past five years which we will now outline. 

\subsection{Perturbations of Schwarzchild Spacetime}

\noindent
 One area which has seen some recent partial results is to prove the Penrose inequality for small perturbations of Schwarzchild spacetime.  This is a physically relevant scenario, since classically in the literature, authors were often interested in small generic perturbations of Schwarzchild spacetime (see, for example, the Penrose singularity theorem) [18, 19].  There are now several results which prove the Penrose inequality for perturbations of Schwarzchild spacetime, albeit with strong restrictions.   Alexakis  used  a  perturbation  argument about a spherically symmetric null hypersurface in Schwarzchild spacetime to prove the null Penrose inequality for  perturbations  of  Schwarzchild  exterior  spacetime  [20].  Recall that the null Penrose inequality is as follows
 
 \[ E_B [ S^{\infty}] \geq \sqrt{\frac{\text{Area} [S']}{16 \pi}}, \tag{41} \]
 
 \noindent
 where $E_B$ is the Bondi energy defined by taking the limit of the Hawking mass
 
 \[ E_B := \lim_{t \rightarrow \infty } m_{\text{Hawk}} [S_t]  . \tag{42}\]
 
 \noindent
 $E_B[S^{\infty}]$ is the Bondi energy in the asymptotically round sphere $S^{\infty}$ on the past-directed outgoing null surface emanating outwards from a perturbation of a suitable null hypersurface.  
 
 Kopinski and Tafel adapted the conformal method from general relativity to prove the Penrose inequality for Schwarzchild initial data under an addition of the axially symmetric traceless exterior curvature, assuming the initial metric to be conformally flat [21].  To be more specific, they take $\Sigma$ to be an initial surface bounded by a $2$-sphere $S$ with radius $M/2$ such that $g_{ij}$ is the flat metric on $\Sigma$ and $K_{ij}$ is a trace-free axially symmetric solution of the momentum constraint which is asymptotically flat such that
 
 \[  K_{ij} = O (r^{-2} ), \: \: \: \: \: \: \: \: \partial_p K_{ij} = O (r^{-3} )  \: \: \: \: \text{as} \: \: \: \: r \rightarrow \infty        .  \tag{43}\]
 
\noindent
Furthermore, one assumes that the Lichnerowicz equation admits a positive solution $\psi$

\[ \Delta \psi = \frac{1}{8} R \psi - \frac{1}{8} K^2 \psi^{-7} \tag{44}   \]

\noindent
which satisfies boundary conditions

\[ n^i \partial_i \psi + \frac{1}{2} h \psi - \frac{1}{4}K_{nn} \psi^{-3} = 0 \: \: \: \: \text{on} \: \: \: \: S,\tag{45a} \]

\[ \psi \rightarrow 1 \: \: \: \: \text{as} \: \: \: \: r \rightarrow \infty . \tag{45b}\]

\noindent
where $h$ is the mean curvature of $S$, $n$ is the unit normal vector, $K_{nn}$ is $K_{ij}$ contracted with two copies of the unit normal vector and $\Delta$ and $R$ are the Laplace operator and the Ricci scalar for $g_{ij}$ .  If one also assumes that $\psi$ can be expanded in powers of a parameter $\epsilon$ which is proportional to some suitable norm of $K_{ij}$, then one has initial data 

\[ g'_{ij}  = \psi^4 g_{ij},  \: \: \: \: K'^i_j =  \psi^{-6} K^i_j    \tag{46}  \]

\noindent
such that a stronger version of the Penrose inequality is satisfied

\[ E^2 - p^2 = \frac{|S|}{16 \pi} + \frac{4 \pi}{| S |} J^2 ,  \tag{47}  \]

\noindent
where $E$, $p$ and $J$ are the total energy, momentum, and angular momentum of the initial data.  If the inner boundary is the one determined by the condition for the conformal factor $\psi$, then this Penrose inequality is satisfied as far as second order in $\epsilon$.  On the other hand, if one takes the initial data to be generalized Bowen-York initial data

\[K_{rr} = \frac{a}{r^3} + 3 \Bigg( \frac{\overline{p}}{r^4} + \frac{p}{r^2} \Bigg) \cos \theta,  \: \: \: \: K^{\theta}_{\theta} = K^{\phi}_{\phi} = - \frac{1}{2}K_{rr}   , \tag{48a}\]

\[ K_{r \theta} =  \frac{3}{2} \Bigg( \frac{\overline{p}}{r^3} - \frac{p}{r} \Bigg) \sin \theta , \: \: \: \:K_{\phi r} = \frac{3 J}{r^2} (z^2 -1) , \: \: \: \: K_{\phi \theta} = 0 ,  \tag{48b}\]

\noindent
 where $a$, $p$, $\overline{p}$ and $J$ are constants, then the Penrose inequality (47) is satisfied up to fourth order in $\epsilon$.  The inequality is saturated for the Schwarzchild metric when $p = \overline{p} = J = 0$.  
 
This result was later extended using similar methods to non-maximal perturbations of Schwarzchild initial data [22], and in this case the initial data satisfy the same Penrose inequality up to second order in $\epsilon$.  This seems like quite a restrictive result, but it is promising as it probably analogous to similar results which were established in the direction of the positive mass theorem before the full result was finally proved by Schoen and Yau (the theorem having first been proved for axially symmetric initial data and for second order perturbations away from flat initial data) [23].  However, the inner boundary in the result for nonmaximal perturbations is a somewhat weak object, being determined by the boundary condition for the conformal factor.  It might be desirable to at least show the same result when the inner boundary is a genuine outermost marginally outer trapped surface.  Finally, Roszkowski and Malec used different techniques to prove the Penrose inequality for Schwarzchild spacetime with linear axial perturbations [24].  Roszkowski and Malec showed that linear axially symmetric perturbations of maximal slices of Schwarzchild spacetime satisfying the Regge-Wheeler gauge condition satisfy the Penrose inequality for apparent horizons up to terms which are linear in a smallness parameter $\epsilon$.  A few assumptions are needed for this argument, including assumption of the existence of the inverse mean curvature flow extended outwards from the outermost apparent horizon plus conditions on the leaves into which the Cauchy hypersurface is assumed to be foliated.  Jezierski had in fact already used different methods somewhat earlier to prove the Penrose inequality for apparent horizons in the case of nonlinear perturbations of the initial data about the Reissner-Nordstrom metric for a spherically symmetric charged black hole [25].

These results all involve some strong restrictions, typically that the perturbations are conformally flat or restrictions on the foliation involved for the Cauchy hypersurface.  It is not clear at present how to obtain stronger results for perturbations of the Schwarzchild metric.  An idea which is very appealing is to take one of the couplings of the generalized Jang equation to the IMCF or the CFM (or even another geometric flow) and then establish an existence and regularity theory for the resulting system in the simplified case where the black hole spacetime in question is a perturbation of Schwarzchild spacetime.  For example, one could take an approach based on a suitable implicit function for Banach spaces to prove the existence of the solution of the generalized Jang equation coupled to the conformal flow of metrics as outlined above for the usual constant $t$ slice of Schwarzchild spacetime.  Unfortunately, it seems at present that even the linearised system is too difficult to analyse, at least with current PDE techniques and a priori estimates.  The method might still be viable for certain specific perturbations of Schwarzchild initial data.

\subsection{Spinor Approaches to the Penrose Inequality}

\noindent
  An alternative approach to the method via PDE analysis is to adapt Witten's original proof with spinors to the case of black hole spacetimes.  Spinor methods have already been successfully employed to extend the positive mass theorem to black hole spacetimes [26].  Herzlich also employed spinor methods to prove a Penrose-type inequality for the mass of an asymptotically flat Riemmanian $3$-manifold with an inner minimal $2$-sphere and non-negative scalar curvature, showing that the total mass is bounded from below by an expression which involves the area of the minimal sphere and a normalized Sobolev ratio [27]
  
  \[ m \geq \frac{1}{2} \frac{\sigma}{1 + \sigma} \sqrt{\frac{\text{Area} ( \partial M) }{\pi}}  , \tag{49}\]
  
  \noindent
  where $\sigma$ is a dimensionless quantify defined as
  
  \[ \sigma = \sqrt{\frac{\text{Area} ( \partial M) }{\pi}}  \inf_{f \in C^{\infty}_c}  \frac{||df||^2_{L^2 (M)}}{||f||^2_{L^2(\partial M)}}. \tag{50}\]
  
  \noindent
Wang also used a spinorial method to define a static quasi-local mass with reference to Schwarzchild space and prove a certain localized Penrose-type inequality: namely, take $\Omega$ to be a compact domain in a maximal asymptotically flat initial data $(M,g)$ with non-negative scalar curvature and the boundary  is a disjoint of two pieces $\Sigma$ and $\Sigma_H$ such that $\Sigma$ has positive mean curvature $H$ and $\Sigma_H$ is a minimal hypersurface with no other closed minimal hypersurfaces in $M$.  If $\Sigma$ is isometric to a closed hypersurface in the spatial Schwarzchild metric with mass $m$ and $\Sigma_H$ is isometric to a hypersurface in the spatial Schwarzchild metric with non-negative mean curvature, then one has

\[ m + \frac{1}{16 \pi} \int_{\Sigma} V \Bigg( \frac{(v H_0)^2}{h} - H \Bigg) \: d \Sigma \geq \sqrt{\frac{| \Sigma_H|}{16 \pi}}  ,\tag{51}\]

\noindent
where $H_0$ is the mean curvature of the surface $\Sigma'_0$ which is conformal to $\Sigma$ in Euclidean space and $V$ is the static potential of Schwarzchild spacetime.  In the proof of this result, note that the spinors are defined when the manifold has a boundary with two components [28].  
  
This raises an intriguing question as to whether the full classical Penrose inequality can be proved with spinors.  In general, a spinor proof would be conceptually clearer and more physically motivated than a proof via geometric analysis.  It is certainly possible to obtain some bound on the mass using the existence of solutions to a spinor equation, but the inequality will depend on the boundary conditions in a restrictive way.  The original argument of Witten used an integral identity for a spinor field $\kappa_A$ over a $3$-dimensional hypersurface $S$.  In the spinor formalism, the troublesome part of the integral for the bulk can be eliminated if $\kappa_A$ solves the Sen-Witten-Dirac equations

\[ \mathcal{D}_{A'}^B \kappa_B = 0 , \tag{52}\]

\noindent
where $\mathcal{D}_{A A'}$ is the derivative defined via 

\[ \mathcal{D}_{A A'} \kappa_C = T_{A A'}^{B B'} \nabla_{B B'} \kappa_C  , \tag{53}\]

\noindent
A key problem with using the Sen-Witten-Dirac equation directly in the black hole setting is that it is a first-order elliptic PDE and so cannot be used to specify all the components of $\kappa_A$ as necessary via suitable asymptotic boundary conditions.  The new approach proposed by Kopinski and Valiente Kroon is to instead study geometric inequalities with the ADM mass but analysing a different equation called the approximate twistor equation [29].  The formalism which they use to do this is that of $1 + 1 + 2$ space spinors: for example, if $\tau^{A A'} $ denotes the spinor counterpart of the timelike normal vector to the hypersurface $S$, then this spinor can be used to write an $SU(2, \mathbb{C})$ version of the above derivative.  

Rather than being first-order, the approximate twistor equation is a second-order elliptic PDE for a Weyl spinor $\kappa_A$ [30].  Formally, this equation can be written as

\[ \textbf{L} (\kappa_A) = \mathcal{D}^{BC} \mathcal{D}_{(AB} \kappa_{C)} - \Omega_{A}^{BCD} \mathcal{D}_{BC} \kappa_{D} = 0 , \tag{54} \]

\noindent
where $\Omega_{ABCD} = K_{(ABC)D}$ and $K_{ABCD}$ corresponds to the extrinsic curvature of a hypersurface orthogonal to $\tau^{A A'}$.  Here $\textbf{L}$ is a composition operator $\textbf{T}^* \circ \textbf{T}$, where $\textbf{T}$ is the spatial twistor operator which maps from the space of valence 1 symmetric spinors over the hypersurface to the space of valence 3 symmetric spinors over the same hypersurface, and $\textbf{T}^*$ is the formal adjoint.  By coupling the approximate twistor equation to a suitable transverse boundary condition, one obtains an elliptic boundary value problem. Considering Green's identity for the approximate twistor operator, Kopinski and Valiente Kroon construct solutions to the boundary value problem and show that for a marginally outer trapped surface defined via suitable spinor analogues of null expansions, one has a so-called master inequality

\[ 4 \pi m  \geq \frac{\kappa}{\sqrt{2}} \textbf{H} [ \phi_A, \overline{\phi}_A ]   ,\tag{55} \]

\noindent
where $\kappa = 8 \pi G / c^4$ and $\textbf{H}$ is the Nester-Witten functional over the marginally outer trapped surface evaluated for a smooth, freely specifiable spinor field $\phi_A$ over the boundary of the hypersurface $S$ (given two spinors $\kappa_A$ and $\omega_A$).  If the freely specifiable spinor $\phi_A$ can be chosen such that $\textbf{H} [ \kappa_A, \omega_B]$ is non-negative, one would immediately have a bound on the ADM mass of a black hole spacetime.  The  result already implies a new proof of the positive mass theorem for black hole spacetimes if we choose the free spinor $\phi_A$ to vanish, but more importantly one can use this master inequality to start deriving geometric inequalities for the ADM mass (possibly even the classical Penrose inequality).  As an example, Kopinski and Valiente Kroon show that if $\phi_A$ is chosen to be an eigenspinor of the $2$-dimensional Dirac operator, an inequality is obtained in the case of a marginally outer trapped surface as follows

\[ 4 \pi \geq  \frac{\sqrt{2}}{2} ( \min_{\partial S} h ) | \partial S |, \tag{56} \]

\noindent
where $h$ is the mean curvature of $S$.  Similarly, they demonstrate that there is a relation between the mass and the spinor counterpart of the variation of the area of a marginally outer trapped surface

\[ 4 \pi m \geq 2 \sqrt{2} \delta_v | \partial S|  . \tag{57}\]

\subsection{Counterexamples to the Penrose Inequality}

\noindent
There has been some interesting recent work on construction of counterexamples to certain versions of the Penrose inequality, with the most famous being perhaps the Carrasco-Mars counterexample [31].  Bray and Khuri had previously formulated a new version of the Penrose inequality in the context of their work on the PDE approach to the Penrose inequality.  This version (which could be called the generalized Penrose inequality) is stated in terms of new types of bounding surface called  generalized trapped surfaces and generalized apparent horizons.  The former obeys the equation $p \leq |q|$ and the latter obeys $p = q$, where $p$ is the mean curvature of a closed surface $S$ embedded in the initial data set and $q$ is the trace of the second fundamental form $K_{ij}$ after it has been pulled back onto $S$.  This definitions are more general than the usual ones, where in terms of the initial data set $(\Sigma, g, K)$ a surface is said to be weakly outer trapped if $p+ q \leq 0$ and marginally outer trapped if $p +q =0$.  The usual concept of an apparent horizon was defined earlier.  Bray and Khuri show that an asymptotically flat initial data set which contains a generalized trapped surface also contains a unique outermost generalized apparent horizon $S_{\text{out}}$.  In terms of this apparent horizon, the generalized Penrose inequality is as follows

\[ M \geq \sqrt{\frac{|S_{\text{out}}|}{16 \pi}}  \tag{58} \]

\noindent
Interestingly, it is possible to construct a counterexample to this version of the conjecture.  The geometric construction is slightly intricate, but essentially one takes the Kruskal spacetime with positive $M$ and shows explicitly that it contains asympotically flat spacelike hypersurfaces with an outermost generalized apparent horizon such that

\[ |S_{\text{out}} | > 16 \pi^2  .\tag{59}\]

This counterexample is sometimes mistakenly believed to be a counterexample to the original classical inequality.  However, as Carrasco and Mars note themselves, the existence of this counterexample only invalidates the use of generalized apparent horizons, and not their general approach via the generalized Jang equation.  In fact, the existence of a counterexample to the generalized Penrose inequality with generalized apparent horizon as the main geometric object is perhaps not as surprising as it seems, since Ben-Dov had already found a counterexample for the Penrose inequality formulated in terms of marginally outer trapped surfaces (a setting which is less general than the one considered by Bray and Khuri) [32].  Ben-Dov constructed an elaborate counterexample in the form of a spacetime composed of four different regions (two of the regions being Schwarzchild spacetimes and two of them being closed dust-filled Robertson-Walker spacetimes).  This example violates a version of the Penrose inequality 

\[ M \geq  \sqrt{\frac{A}{16 \pi}} , \tag{60}\]

\noindent
where $M$ is the total mass of the spacetime and $A$ is the area of the marginally outer trapped surface (note that Ben-Dov refers to marginally outer trapped surfaces as 'apparent horizons').  Strictly speaking, this is only a version of the time-symmetric Riemannian Penrose inequality.  One could write the general version by replacing the area of the marginally outer trapped surface with the minimum area enclosure of the marginally outer trapped surface, but there is no physical reason to expect non-existence of counterexamples to this version.  The fact that a counterexample exists to the Riemannian Penrose inequality for marginally outer trapped surfaces is not intrinsically surprising in itself [33], but the counterexample is physically interesting because in many ways it parallels the original construction of Penrose in terms of a collapsing shell of null dust [18].

The Penrose inequality is closely allied to the cosmic censorship hypothesis.  A proof of the Penrose inequality would give strong indirect evidence for the hypothesis, since cosmic censorship was used in the original derivation of the inequality by Penrose [18].  We do not have a completely satisfactory formal statement of this hypothesis, but at present the weak cosmic censorship hypothesis as formulated by Penrose [10, 19] and developed by Geroch and Horowitz [34] states that a globally hyberbolic spacetime which asymptotes to Minkowski spacetime at spatial infinity cannot develop a singularity which is causally connected with future null infinity (known as a naked singularity).  In other words, if we formulate general relativity in such a way that it still makes sense to talk about casuality and determinism, a singularity can only exist if it is behind an event horizon where it does not cause the evolution of outside trajectories to deviate from predictability.  There do exist counterexamples to the cosmic censorship hypothesis, but they are not strict counterexamples because they are unphysical and only violate cosmic censorship 'formally'.  A counterexample to the cosmic censorship hypothesis was found by Roberts [35]

\[ ds^2 = - (1+ 2 \sigma) dv^2  + 2 dv dr + (r^2 - 2 \sigma r v) d \Sigma^2  ,\tag{61a}\]

\[ \phi = \frac{1}{2} \log \Bigg( 1 - \frac{2 \sigma v}{r} \Bigg),\tag{61b} \]

\noindent
where $\sigma > -1/2$.  This is however a non-static solution to the massless scalar Einstein equations rather than the classical Einstein equations

\[ 8 \pi T_{ab} = 2 \phi_a \phi_b - g_{ab} \phi_c \phi^c .\tag{62}\]

\noindent
These equations are related to the vacuum Brans-Dicke field equations where the gravitational interaction is mediated by a scalar field $\phi$ as well as a tensor field.  The standard results and singularity theorems also hold in Brans-Dicke theory since gravitation remains attractive in this setting, but the theory is not widely believed to be a viable competitor with general relativity even if it has not yet contradicted physical observations [36, 37].  Also note that the solution is spherically symmetric (as was the counterexample to the Penrose inequality of Ben-Dov), a somewhat special assumption which can be ruled out of realistic models by certain hypotheses.  Similarly, a plausible counterexample to the weak cosmic censorship hypothesis has recently been proposed for four-dimensional gravity with asymptotically flat boundary conditions.  However, this is in the case of Einstein-Scalar theory where the Einstein-Hilbert action is coupled to a massive scalar field $\psi$ such that the Einstein equations take the same form but the energy momentum tensor is now 

\[  T_{ab} = 2 \nabla_a \psi \nabla_b \psi - g_{ab} \nabla^c \psi \nabla_c \psi - \mu^2 \psi^2 g_{ab},\tag{63} \]

\noindent
where $\mu$ is the mass of the scalar field [38].  At the time of writing, this example is plausible due to numerical support, but is not known to be a counterexample.  

Given the existence of certain known or plausible counterexamples to the cosmic censorship hypothesis, one might also ask if it is possible to have unphysical or unrealistic violations of the classical Penrose inequality such that one constructs some strange or intricate black hole spacetime with no regard to realism and then shows that its total mass $M$ contradicts the inequality.  This is also an open question, since very little progress has been made on cosmic censorship since the original paper of Penrose [19].  At present, there is not yet any evidence for a counterexample to the classical inequality even with some strange matter, so it looks for the moment as if the inequality is absolutely not violated even under unphysical circumstances [39].  This is a possibly subtle point which does not seem to have been discussed much in the literature, but this could simply be because the classical Penrose inequality has a relatively clear statement in comparison with the cosmic censorship hypothesis.

\subsection{Extended Penrose-type Inequalities}

\noindent
The stronger version of the Penrose inequality proved by Kopinski and Tafel incorporates angular momentum.  Using arguments similar to those of Penrose [40], it can be shown heuristically that the corresponding Penrose-like inequality with angular momentum for an axially symmetric initial data $\Sigma$ should be 

\[ M^2 \geq \frac{A}{16 \pi} + \frac{4 \pi J^2}{A} ,\tag{64}\]

\noindent
with saturation occurring only when the initial data embed into the Kerr spacetime for a rotating black hole.  The assumption of axial symmetry is necessary in order to define a quasi-local notion of angular momentum.  In general, it has taken even longer to start arriving at rigorous results about geometric inequalities in relativity which involve angular momentum because the definition of angular momentum is more subtle than the definition of mass [41].  In this direction, Anglada used the monotonicity of the Geroch energy along the inverse mean curvature flow to prove the following Penrose inequality with angular momentum

\[ m_{ADM}^2 \geq \frac{A}{16 \pi} + \frac{4 \pi J^2}{2 R^2} ,\tag{65}\]

\noindent
where $m_{ADM}$ is the ADM mass, $A$ is the area of an apparent horizon which is also a compact connected minimal surface, and $R$ is a measure of size defined via the norm of the axial Killing vector [40].  This result also assumes maximal initial data in order to have non-negative scalar curvature.  Anglada has recently used similar techniques and the monotonicity properties of the Hawking rather than the Geroch mass to prove the same Penrose inequality for axially symmetric initial data such that the boundary is a compact connected outermost apparent horizon, as long as one assumes existence of a smooth inverse mean curvature flow of surfaces which starts with spherical topology [42].

Extensions to non-maximal initial data were implemented by Jaracz and Khuri, who showed that if one starts with a complete, axisymmetric, asympototically flat initial data set $(M,g,k,E,B)$ for the Einstein-Maxwell equations which is devoid of apparent horizons and satisfies the charged dominant energy condition $\mu_{\text{EM} \geq |J_{\text{EM}} |},$ then the following inequality holds

\[ \mathcal{E}  \geq \frac{Q^2}{2 \overline{\mathcal{R}}} + \frac{G}{2 c^2} \frac{\mathcal{J}^2}{\overline{\mathcal{R}} \: \overline{\mathcal{R}}^2_c}  ,\tag{66} \]

\noindent
where $G$ is the gravitational constant, $c$ is the speed of light, $Q$ is the total charge of a connected open subset $\Omega \subset \Sigma$ with compact closure and smooth boundary known as a body, $\mathcal{J}$ is the angular momentum of the body, $\overline{\mathcal{R}}$ is the radius of the smallest sphere which encloses the body, $\mathcal{E}$ is the total energy of such a sphere,  $\overline{\mathcal{R}}_c$ is the special radius used by Anglada known as the circumference radius and $E$ and $B$ are the electric and magnetic fields induced on $M$ [43, 44].  The theorem also assumes that there is no charge density or momentum density in the direction of axisymmetry outside of the body and that the Jang-IMCF system admits solutions.  This generalises the results of Anglada to the non-maximal setting.  This is a Bekenstein-type inequality but a similar Penrose-type inequality can also be established for the same type of initial data set and the same energy codition.  In this setting, we must now assume that the outermost apparent horizon boundary only has one component and that the Jang-IMCF system has well-behaved solutions which do not exhibit the jumping behaviour which is typical for the inverse mean curvature flow:

\[ \mathcal{E}^2 \geq \Bigg( \frac{c^4}{G} \sqrt{\frac{|\partial M |}{16 \pi}} + \sqrt{\frac{\pi}{|\partial M |}} Q^2 \Bigg)^2 + \frac{c^2 \mathcal{J}^2}{4 \overline{\mathcal{R}}_c^2} . \tag{67}\]

\noindent
In the case of spherically symmetric initial data, it has been shown rigorously that such solutions to the Jang-IMCF system necessarily exist [14].  

Motivated by the analogous results for the ADM mass, Alaee, Khuri and Yau have recently established a range of Penrose-type inequalities for different types of quasi-local mass [45].  The first of these is stated in terms of the Brown-York mass, which is defined as [46]

\[ m_{\text{BY}} = \frac{1}{8 \pi} \int_{\Sigma} (H-H_0) \: d A_{\sigma}   ,\tag{68}\]

\noindent
where $H$ is the mean curvature of a spacelike $2$-surface $\Sigma$ which bounds a compact spacelike hypersurface $\Omega$ in a spacetime equipped with a time orientation, $H_0$ is the mean curvature of an isometric embedding of $\Sigma$ into $\mathbb{R}^3$ and $\sigma$ is the induced metric on $\Sigma$.  The corresponding Penrose-type inequality with the Brown-York mass includes charge and angular momentum.  If one starts with the charged dominant energy condition and a maximal axisymmetric spacelike hypersurface $(\Omega, g,k)$ such that $\Omega$ is either compact or has asymptotically cylindrical ends with limiting cross-sections $\Sigma_{\infty}$ along with the assumption that the generalized inner boundary formed by the union of $\Sigma_{\infty}$ with pieces of $\Omega$ which only contain apparent horizon components $\Sigma_h$ is strictly outer minimizing and axisymmetric and the assumption that the single component $\Sigma$ is a mean convex surface with positive Gaussian curvature, then one has

\[ m_{\text{BY}} ( \Sigma) \geq \frac{1}{2} \Bigg( \sum_{i=1}^I \sqrt{(Q^i_h)^4 + 4 (\mathcal{J}_h^i)^2} + \sum_{j=1}^J \sqrt{(Q^j_{\infty})^4 + 4 (\mathcal{J}_{\infty}^j)^2} \Bigg)^{1/2}. \tag{69} \]

\noindent
It is also required that $\Sigma_h$ is the only compact minimal surface and that the electric and magnetic fields be divergence-free.  Furthermore, if the generalized inner boundary $\Sigma_{*}$ only has one component, 

\[  m_{\text{BY}} ( \Sigma)^2 \geq \Bigg( \sqrt{\frac{|\Sigma_{*}|}{16 \pi}} + \alpha^2 \sqrt{\frac{\pi}{| \Sigma^* | }} Q^2 \Bigg)^2 + \frac{(\alpha \mathcal{R}_* \mathcal{R}_c^{-1} )^2}{2} \frac{4 \pi \mathcal{J}^2}{| \Sigma_* |} , \tag{70}\]

\noindent
where

\[ \mathcal{R}_* = \sqrt{\frac{|\Sigma_* |}{16 \pi}}, \:\:\:\:\:  \alpha^2 = 1 - \sqrt{\frac{|S_0|}{|S_{t_0}|}}, \tag{71} \] 

\noindent
where $\{ S_t \}_{t=0}^{t_0} $ is the unique inverse mean curvature flow defined for all extensions of an outer minimizing $\Omega$ known as the indigeneous inverse mean curvature flow and $t_0$ corresponds to the leaf of greatest area contained in $\Omega$.  The generalized inner boundary is said to be outer minimizing if the area of all possible enclosing surfaces in $\Omega$ is not less than $|\Omega_h| + |\Omega_{\infty}|$.  This inequality also implies a Bekenstein-type bound.  Note that $\mathcal{J}$ is the Brown-York angular momentum which is defined as

\[ \mathcal{J}_{\text{BY}} = \int_{\Sigma} p( \eta, \nu ) \: d A_{\sigma}   ,\tag{72}\]

\noindent
where $\nu$ is the spacelike unit normal to $\Sigma$ pointing out of $\Omega$, $p$ is the momentum tensor and $\eta$ is the generator of the $U(1)$ symmetry under which all the relevant quantities are invariant.  

The Brown-York mass has some deficiencies which has led to alternative definitions of quasi-local mass known as the Liu-Yau and the Wang-Yau mass [47]:  

\[ m_{\text{LY}}(\Sigma) =  \frac{1}{8 \pi}  \int_{\Sigma}  \Bigg( | \vec{H}_0 | - | \vec{H} | \Bigg) \: dA_{\sigma}   ,\tag{73} \]

\[ m_{\text{WY}} (\Sigma)  = \frac{1}{8 \pi} \int_{\Sigma} \frak{p} \: d A_{\sigma}  ,\tag{74}\]

\noindent
where $\vec{H}_0 = H_0 \nu_0$ is the mean curvature vector of the isometric embedding of $\Sigma$ into the Euclidean time hypersurface contained in Minkowski spacetime, $\vec{H}$ is the spacetime mean curvature vector and $\frak{p}$ is defined to be

\[ \frak{p} = \frac{\sqrt{| \vec{H}_0 |^2 + \frac{(\Delta \tau)^2}{1 +|\nabla \tau |^2 }} - \sqrt{| \vec{H} |^2 + \frac{(\Delta \tau)^2}{1 +|\nabla \tau |^2 }}}{\sqrt{1 + |\nabla \tau|^2}}    ,\tag{75}\]

\noindent
for a time function $\tau = - \langle T_0, \frak{i} \rangle$ defined for a future timelike unit Killing field on Minkowski spacetime and $\frak{i}$ an isometric embedding from $\Sigma$ into Minkowski spacetime.  Note that the definition for the Wang-Yau mass is for the case of an optimal isometric embedding.  Similarly, one can define suitable Liu-Yau and Chen-Wang-Yau quasi-local angular momenta [48]:

\[ \mathcal{J}_{\text{LY}}  = \frac{1}{8 \pi} \int_{\Sigma} \textbf{j} (\eta^T) \: d A_{\sigma}  ,\tag{76}\]

\[ \mathcal{J}_{\text{CWY}} = \int_{\Sigma} \Bigg( \frak{p} \langle \eta, T_0 \rangle + \frak{j} (\eta^T)  \Bigg) \: d A_{\sigma}   ,\tag{77}\]

\noindent
where $\eta^T$ is the tangential part of an axisymmetric Killing field for the embedding of $\Sigma$ into $\mathbb{R}^3$, $\eta$ is a Killing field which generates the axisymmetry in Minkowski spacetime and $\frak{j}$ is defined to be

\[ \frak{j} = \nabla \tau - \nabla \sinh^{-1}  \Bigg( \frac{\frak{p} \Delta \tau}{| \vec{H}_0 | |\vec{H} |} \Bigg) + \alpha_{\vec{H}_0} - \alpha_{\vec{H}}\tag{78}  \]

\noindent
for $\alpha$ the connection $1$-form of the normal bundle in a suitable gauge.  Finally, $\textbf{j}$ is defined to be

\[ \textbf{j} =  * d \omega   , \tag{79}\]

\noindent
for the Hodge star $*$ associated to the induced metric $\sigma$ and $\omega$ being one of the two functions used to define the Hodge decomposition of the Brown-York momentum surface density $1$-form.

Penrose-type inequalities can be stated both in terms of the Liu-Yau and the Wang-Yau mass [45].  Start with an initial data set $(\Omega, g, k)$ for the Einstein equations satisfying the dominant energy condition, it is also assumed that $\Omega$ is compact with a boundary which is composed of a disjoint union of a single component $\Sigma$ and a single component apparent horizon $\Sigma_h$, that there are no other closed apparent horizons, and that $\Sigma $ is a non-trapped surface with positive Gaussian curvature.  If all of these hold, there is a non-zero constant $\gamma$ which does not depend on the area of the horizon such that

\[ m_{\text{LY}} (\Sigma) = \frac{\gamma}{1 + \gamma} \sqrt{\frac{| \Sigma_h |}{4 \pi}}  . \tag{80} \]

\noindent
This can be improved when the horizon has a single component: starting with an an axisymmetric initial data set $(\Omega, g,k, E,B)$ for the Einstein-Maxwell equations such that the Killing field $\eta$ satisfies

  \[ \eta \: \wedge \: d \eta = 0  \]

\noindent
and such that $J(\eta) =0$ and $\mu_{\text{EM}} \geq |J|$, the same conditions are assumed as before but now $\Omega$ also has an axisymmetric boundary.  If all of these hold, there are then two non-zero constants $\lambda$ and $\gamma$ which do not depend on the area of the horizon such that 

\[ m_{\text{LY}}(\Sigma)^2 \geq \Bigg( \frac{\gamma}{1 + \gamma} \sqrt{\frac{| \Sigma_h |}{4 \pi}} + \lambda \sqrt{\frac{\pi}{|\Sigma_h |}} Q^2 \Bigg)^2  + \frac{\lambda \gamma}{1 + \gamma} \frac{8 \pi^2 \mathcal{J}^2}{\mathcal{C}^2} , \tag{81}\]

\noindent
where $\mathcal{C}$ is the largest circumference of all $\eta$-orbits in $\Omega$.  As before, this Penrose-type inequality also implies a Bekenstein-type bound.  The same inequality holds without any symmetry assumptions if the contribution due to angular momentum is set to zero.  Similar Penrose-type inequalities can also be stated for the Wang-Yau mass with the geometric requirements taking a similar form, but one now requires the notion of a Wang-Yau data set $(\Sigma, \sigma, |\vec{H} |, \alpha)$ with an optimal isometric embedding.  Finally, they demonstrated that these Penrose-like inequalities can be extended to quasi-local masses which are defined via a static reference spacetime apart from Minkowski spacetime.  As an example, they show that a Penrose-type inequality holds for the static Liu-Yau mass when the reference spacetime is Schwarzchild spacetime, extending the previous result of Lu and Miao who proved such a Penrose inequality for the static Brown-York mass [49].

\section*{References}

\noindent
[1] Galloway G 2017 Topology and general relativity.  http://www.math.miami.edu/~galloway/ESI2017.pdf

\noindent
[2] Witten E 2020 Light Rays, Singularities, and All That.  Rev. Mod. Phys. 92, 045004

\noindent
[3] Lee JM 1997 Riemannian Manifolds: An Introduction to Curvature (Springer)

\noindent
[4]  Galloway G and Schoen R 2006 A generalization of Hawking's black hole topology theorem to higher dimensions.  Comm. Math. Phys., 266:2

\noindent
[5] Choquet-Bruhat Y 2015 Introduction to General Relativity, Black Holes, and Cosmology (Oxford University Press)

\noindent
[6] Bunting G and  Masood-ul-Alam A 1987 Nonexistence of multiple black holes in asymptotically Euclidean static vacuum space-time. Gen. Relativity Gravitation, 19:2 

\noindent
[7] Schoen R and Yau S-T 1979 On the proof of the positive mass conjecture in general relativity.  Comm. Math. Phys., 65:45-76

\noindent
[8] Schoen R and Yau S-T 1981 Proof of the positive mass theorem II.  Comm. Math. Phys., 79:231-260 

\noindent
[9] Witten E 1981 A new proof of the positive energy theorem.  Comm. Math. Phys., 80:3 

\noindent
[10] Penrose R 1973 Naked singularities.  Ann, N.Y. Acad. Sci. 224, 125-134.

\noindent
[11] Huisken G and Ilmanen T 2001 The inverse mean curvature flow and the Riemannian Penrose inequality.  Journal of Differential Geometry, 59:3 

\noindent
[12] Bray H 2001 Proof of the Riemannian Penrose inequality using the positive mass theorem.  Journal of Differential Geometry, 59:2 

\noindent
[13] Bray H 2014 Time Flat Surfaces and the Monotonicity of the Spacetime Hawking Mass.  Comm. Math. Phys., 331(1): 285-307

\noindent
[14] Bray HL and Khuri M 2010 A Jang Equation Approach to the Penrose Inequality.  Discrete Contin. Dyn. Syst., 27 741-766. 

\noindent
[15] Bray H and Khuri M 2011 PDEs which imply the Penrose conjecture.   Asian J. Math., 15:4

\noindent
[16] Han Q and Khuri M 2013 Existence and Blow-up Behaviour for Solutions of the Generalized Jang Equation.  Comm. Partial Differential Equations. 38 2199-2237.

\noindent
[17] Han Q and Khuri M 2018 The Conformal Flow of Metrics and the General Penrose Inequality.  Adv. Math. Phys., Vol. 2018, Art. ID 7390148.

\noindent
[18] Penrose R 1969 Gravitational collapse: The role of general relativity.  Nuovo Cimento. Rivista Serie 1: 252-276

\noindent
[19] Penrose R 1999 The question of cosmic censorship.  J. Astrophys. Astron. 20, 233


 \noindent
 [20] Alexakis S 2015 The Penrose inequality on perturbations of the Schwarzschild exterior.  arXiv:1506.06400v1.

\noindent
[21]  Kopinski J and Tafel J 2019 The Penrose inequality for perturbations of the Schwarzschild initial data.  Class. Quantum Grav. 37 015012.

\noindent
[22] Kopinski J and Tafel J (2020) The Penrose inequality for nonmaximal perturbations of the Schwarzschild initial data.  Class. Quantum Grav. 37 (10).

\noindent
[23] Brill D and Deser S 1968 Variational methods and positive energy in general relativity.  Ann. Phys. 50, 548

\noindent
[24] Roszkowski K and Malec E 2005 The Penrose inequality in perturbed Schwarzschild geometries.  Acta Phys. Polon. B 36 2931.

\noindent
[25] Jezierski J 1994 Perturbation of initial data for spherically symmetric charged black hole and Penrose conjecture.  Acta Physica Polonica Series B, 25(10):1413 

\noindent
[26] Herzlich M 1998 “The positive mass theorem for black holes revisited” Journal of Geom. and Phys. 26
97-111

\noindent
[27] Herzlich M 1997 A Penrose-like inequality for the mass of Riemannian asymptotically flat manifolds.  Comm. Math. Phys. 188:121

\noindent
[28] Wang Q 2019 A localized Penrose inequality and quasi-local mass.  Int. Jour. Math. (30) 131940008

\noindent
[29] Kopinski J and Valiente Kroon JA 2020 A new spinorial approach to mass inequalities for black holes in General Relativity.   arXiv:2010.03299v1

\noindent
[30] Backdahl T and Valiente Kroon JA 2011 Approximate twistors and positive mass.  Class. Quantum. Grav., 28:075010

\noindent
[31] Carrasco A and Mars M 2010 A counterexample to a recent version of the Penrose conjecture.  Class. Quantum Grav. 27, 062001

\noindent
[32] Ben-Dov I 2004 Penrose inequality and apparent horizons.  Phys. Rev. D 70, 124031

\noindent
[33] Bray H and Chrusciel P 2004 The Penrose Inequality, in The Einstein Equations and the Large Scale Behaviour of Gravitational Fields (50 years of the Cauchy problem in general relativity), ed. by Friedrich H and Chrusciel P (Birkhaeuser)

\noindent
[34] Geroch R and Horowitz 1979 Global structure of spacetimes in General Relativity: An
Einstein Centenary Survey, edited by Hawking SW and Israel W (Cambridge University Press).

\noindent
[35] Roberts MD 1989 Scalar field counterexamples to the cosmic censorship hypothesis.  Gen. Rel. Gravitation 21, 907-939

\noindent
[36] Hawking SW and Penrose R 1970 The singularities of gravitational collapse and cosmology.  Proc. Roy. Soc. Lond. A 314, 529-548

\noindent
[37] Brans CH and Dicke RH 1961 Mach's Principle and a Relativistic Theory of Gravitation.  Phys. Rev. 124(3), 925-935

\noindent
[38] Eperon FC, Ganchev B and Santos JE 2020 Plausible scenario for a generic violation of the weak cosmic censorship conjecture in asymptotically flat four dimensions.  Phys. Rev. D 101, 041502


\noindent
[39] Senovilla JMM and Garfinkle D 2015 The 1965 Penrose singularity theorem.  Class. Quantum Grav. 32, 124008

\noindent
[40] Anglada P 2018 Penrose-like inequality with angular momentum for minimal surfaces.  Class. Quant. Grav. 35(4):045018

\noindent
[41] Arnowitt R, Deser S and Misner CW 1962 The dynamics of general relativity in Gravitation: An Introduction to Current Research, ed. by Witten L (Wiley)

\noindent
[42] Anglada P 2020 Comments on Penrose inequality with angular momentum for outermost apparent horizons.  Class. Quant. Grav. 37, 065023

\noindent
[43] Anglada P, Gabach-Clement M and Ortiz O 2017 Size, angular momentum and mass for objects.  Class. Quant. Grav. 34, 125011

\noindent
[44] Jaracz JS and Khuri MA 2018 Bekenstein bounds, Penrose inequalities, and black hole formation.  Phys. Rev. D 97(12), 124026

\noindent
[45] Alaee A, Khuri M and Yau S-T 2020 Geometric Inequalities for Quasi-Local Masses.  Comm. Math. Phys 378, 467-505

\noindent
[46] Brown J and York J 1993 Quasilocal energy and conserved charges derived from the gravitational action.  Phys. Rev. D 47(4), 1407

\noindent
[47] Liu C-C and Yau S-T 2003 Positivity of quasilocal mass.  Phys. Rev. Lett. 90(23), 231102

\noindent
[48] Chen P-N, Wang MT and Yau S-T 2016 Quasilocal angular momentum and center of mass in general relativity.  Adv. Theor. Math. Phys. 20, 671-682

\noindent
[49] Lu S and Miao P 2017 Minimal hypersurfaces and boundary behaviour of compact manifolds with nonnegative scalar curvature.  J. Diff. Geo 113(3)

\noindent
[50] Mars M 2009 Present status of the Penrose inequality.  Class. Quantum Grav. 26 193001

\end{document}